\documentclass[10pt]{llncs}
\usepackage{amsfonts, latexsym, amsmath}
\usepackage{color} 
\usepackage[latin1]{inputenc}
\usepackage[english]{babel}
\usepackage{fixltx2e}
\usepackage{graphicx}
\usepackage[hyphens]{url}

\usepackage{breakurl}
\usepackage{paralist}
\usepackage{hyperref} 
\usepackage{tikz}
\usetikzlibrary{decorations.markings}

\newtheorem{defin}{Definition}

\newcommand{\txpublish}{\texttt{TX}_{\texttt{PUBLISH}}}

\newcommand{\txrequest}{\texttt{TX}_{\texttt{REQUEST}}}
\newcommand{\txaccept}{\texttt{TX}_{\texttt{ACCEPT}}}

\newcommand{\addr}{\texttt{a}}

\newcommand{\msig}{\texttt{MSIG}}

\newcommand{\opreturn}{\texttt{OP\_RETURN}}

\newcommand{\txrequestdouble}{\texttt{TX}_{\texttt{REQUEST-DOUBLE}}}
\newcommand{\txacceptdouble}{\texttt{TX}_{\texttt{ACCEPT-DOUBLE}}}

\newcommand{\txrevoke}{\texttt{TX}_{\texttt{REVOKE}}}
\newcommand{\fpublish}{\texttt{F}_{\texttt{PUBLISH}}}
\newcommand{\frequest}{\texttt{F}_{\texttt{REQUEST}}}
\newcommand{\faccept}{\texttt{F}_{\texttt{ACCEPT}}}

\newcommand{\frequestdouble}{\texttt{F}_{\texttt{REQUEST-DOUBLE}}}
\newcommand{\facceptdouble}{\texttt{F}_{\texttt{ACCEPT-DOUBLE}}}

\newcommand{\fgeneric}{\texttt{F}_{\texttt{NAME-OF-TRANSACTION}}}

\newcommand{\Prover}{\mathcal{P}}
\newcommand{\Verifier}{\mathcal{V}}
\newcommand{\Hash}{\mathcal{H}}

\newcommand{\IP}{\mathcal{IP}}
\newcommand{\SP}{\mathcal{SP}}
\newcommand{\USR}{\mathcal{USR}}

\makeatletter \g@addto@macro\@floatboxreset\centering \makeatother 

\title{A User-centric System for Verified Identities on the Bitcoin Blockchain}

\author{Daniel Augot \inst{1} \inst{2} \inst{3}, Herv\'e Chabanne \inst{4} \inst{5}, Thomas Chenevier \inst{4}, William George \inst{2} \inst{1} \inst{3}, Laurent Lambert \inst{4}}

\institute{INRIA, Palaiseau, France \\ \email{daniel.augot@inria.fr} 
	\and 
	Laboratoire LIX, \'Ecole Polytechnique \& CNRS UMR 7161, Palaiseau, France \\ \email{wgeorge@lix.polytechnique.fr}  
	\and
	Universit\'e Paris-Saclay, France
	\and
	OT-Morpho, Issy-les-Moulineaux, France \\ \email{herve.chabanne@morpho.com} \\ \email{thomas.chenevier@morpho.com	} \\ \email{laurent.lambert@morpho.com}
	\and
	T\'el\'ecom ParisTech, Paris, France }

\begin{document}

\date{}
\maketitle
\pagestyle{plain}
\begin{abstract}
  
   We present an identity management scheme built into the Bitcoin blockchain, 
   allowing for identities that are as indelible as the blockchain itself. Moreover, we take advantage of Bitcoin's decentralized nature to facilitate a shared control between users and identity providers, allowing users to directly manage their own identities, fluidly coordinating identities from different providers, even as identity providers can revoke identities and impose controls.
  

\end{abstract}



\section{Introduction}



We live in a world where the ways in which a person's identity is being used are increasingly complex. Appropriately handling sensitive personal data, such as medical, financial, and employment data, is subtle and requires care \cite{CanadaPrivacy}. In this context, it is important to employ technical solutions that promote good security practices and that ensure that users have appropriate controls over how their data is being used. There are many \cite{PIMS} who advocate for a decentralized approach in which users directly manage their own identities via personal servers,
 Personal Identity Management
Systems (PIMS). 
Meanwhile, blockchains, most notably Bitcoin
\cite{satoshi}, have provided new models of decentralization. In this
work, we propose a sort of ``light-PIMS,'' to be implemented on the Bitcoin blockchain. The decentralized nature of the blockchain
allows us to create a neutral space where identity issuers and users
share responsibility for users' identities, providing protections and
the capacity for oversight for both parties.


{\bf Related Work.} 
In 2015 MIT Media Labs introduced a system for academic
certificates on the Bitcoin blockchain \cite{MIT}. Taking advantage of
the blockchain's persistence over time, this system gives students a
convenient way of proving that they graduated, see Section \ref{sec:BuildingBlocks}. The Blockstack project \cite{Blockstack} has implemented decentralized versions of PKI and DNS on the Bitcoin network. In \cite{JH}, a decentralized scheme to issue credentials in the absence of a trusted third party is proposed using Bitcoin. This scheme incorporates zero-knowledge protections such as those we will deal in Section \ref{DLREP}. The startup CryptID \cite{CryptID} has proposed a system where encrypted records of fingerprints (along with a password) are stored in the Factom blockchain, which is itself periodically committed to the Bitcoin blockchain, replacing the traditional centralized server in fingerprint scanning identification systems with a more lightweight system.    
We generalize these ideas to permit more flexible user identities
that can contain different
fields of information useful in interacting with diverse service providers.  We further explore the possibilities
enabled by performing these interactions on a blockchain. 
Some architectures propose new, application designed blockchains. For example, the proposal of IDCoins \cite{IDCoins} relies on a custom blockchain in which the proof of work is related to the generation of GPG/PGP keys necessary to create a web of trust. 
The Guardtime KSI blockchain, which forms the base of an electronic records system used in Estonia \cite{eestonia}, \cite{estoniannews}, is a permissioned blockchain.
In \cite{Zyskind} a system is
proposed to store user information such as the GPS data from
their phone in a distributed hash table and then store pointers to
this data and permissions on how it may be used or retrieved on
a blockchain.
The proposition of ChainAnchor \cite{ChainAnchor} even allows to create a semi-permissioned structure that can be placed on top of an existing blockchain such as that of Bitcoin by changing the incentive structure of miners to promote permissioned transactions. For a survey on other proposals that touch on the relationship between blockchains and identity management, see \cite{BlockIDSurvey} and \cite{BlockStartupSurvey}.  

{\bf Our Contribution.} 
We propose an identity management system that will take advantage of the decentralized nature of the Bitcoin blockchain to allow for a balance between the ability for users to manage their own identities and for issuers to establish controls. The different entities of our
proposal communicate via Bitcoin transactions, allowing identity issuers to outsource much of the infrastructure required for this system to the Bitcoin network, which as the most robust, most established blockchain, has strong security properties, most notably, that
miner's work maintains strong integrity of its data. Privacy during
identity verification is ensured thanks to
the attribute-based credentials of Brands \cite{Brands}. While \cite{JH} already proposes using Brands credentials in Bitcoin, their protocol could, in fact, be implemented in any blockchain without major modifications. In contrast, our proposal takes advantage of the specifics of the Bitcoin scripting language to encode identity meaning in Bitcoin syntax. Specifically, 
we build upon the idea of MIT Media Labs \cite{MIT} that revocation can be encoded in terms of the status of a Bitcoin transaction to enable additional mechanisms for issuer oversight which are then enforced by the Bitcoin network. Particularly, in our system an issuer can limit the number of times an identity can be used, see Section \ref{limiteduse}. At the same time, we will see that our system gives a great deal of control to the user over her identity. 

Note that in traditional systems the reconciliation between user control and issuer oversight is problematic; in most systems the identity is generally controlled entirely by an on-line issuer with little input from the user \cite{PIMS}, or alternatively the issuer will sign an identity to be managed by the user, then the issuer will go offline ceding his capacity for oversight (See \cite{DBLP:journals/ieeesp/CamenischLN12} for a further discussion on the advantages and disadvantages of these two models.) 



Compared to a traditional decentralized system, we offer more integrated issuer controls. For example, compare the revocation mechanism discussed in Section \ref{sec:operational} to the challenges encountered using revocation lists in public key infrastructures (PKIs) \cite{revocationlists}. Additionally, we will see that our system has the following advantages compared to centralized systems:
\begin{itemize}
	\item Our system does not require identity providers to be as ``lively'' as they must be in traditional, centralized systems. If an identity provider has placed controls on an identity, such as a limit on the number of times it can be used, 
	then even if an identity issuer has a service interruption, a user can continue to use her identity and these limits will continue to be enforced by the robust, worldwide Bitcoin network. A user can even revoke her own identity without intervention by the issuer.
	
	\item By providing a common space, control over which is shared between the different actors through the mechanisms of the blockchain, we allow users to coordinate several micro-identities, only needing to trust a small portion of their identities to any given identity provider, see Section \ref{subsec:transactioncoordinate}. While a similar coordinate scheme is possible without a blockchain, in practice it is highly impractical for a user to coordinate identities from different identity providers each of whom uses his own distinct formatting and infrastructure.

\end{itemize}

On the other hand, our system has two (potential) drawbacks. First, as authentications are encoded in Bitcoin transactions, this requires paying transaction fees to miners, see Section \ref{cost} for an estimate of these fees. Second, Bitcoin transactions are by their nature public, posing risks to user anonymity. The typical suggestion to ensure (pseudo-)anonymity in Bitcoin is to use each Bitcoin address exactly one time. An analogous idea works here, at the expense of having higher user fees, see Remark \ref{remark: loopingmodels}. Note that users have differing standards
regarding the privacy that they expect in their
interactions. Some users may be willing to sacrifice some anonymity in exchange for lower fees. In fact, some users, such as those that
gladly link their Facebook account to their Instagram account or
their favorite blogs, may even prefer that metadata on their
transactions be tied to them, allowing them to create a digital
presence on which they can build a reputation. See Section \ref{repu} for a proposal on how a reputation system can be built on top of our architecture. A user should be
empowered to make choices regarding how private they want to be.

\section{Background} \label{sec:BuildingBlocks}

In this section we briefly recall some of the existing ideas, in Bitcoin and in the work of Brands \cite{Brands}, upon which our system is built.

\subsection{Bitcoin relevant notions} \label{subsec:Bitcoinsyntax} 
It is a particularity of Bitcoin that all
bitcoins exist in the form of Unspent Transaction Outputs (UTXOs)
\cite{satoshi}, \cite[Chapter 5]{MasteringBitcoin}. Each transaction may have
several inputs, each of which was an output UTXO for some previous 
bitcoin transaction, and it may have several outputs.
Most transaction outputs correspond to a bitcoin address, the hash of the public key
that can spend it or a hash of a script detailing how the coin can 
be reclaimed. (These are called Pay to Public Key Hash P2PKH and Pay to Script 
Hash P2SH outputs respectively.) Particularly, one can create P2SH outputs that can then be spend by an $m$ of $n$ multisig. Also relevant to our work will be $\opreturn$ 
outputs; each such output contains up to $80$ bytes of space in which the sender
of a transaction can store arbitrary information. Note, $\opreturn$ outputs
must have zero bitcoins associated to them; as such, they are provably not usable as
inputs to later transactions. 

 The raw transaction that is broadcast to the nodes 
contains the amount of bitcoin to associate to each output, the script 
permitting validation of each output (P2SH, P2PKH, etc), and the scripts
for each input that satisfy the requirements set up when the
corresponding input UTXO was created, generally including a signature
from a corresponding private key. The hash of this raw transaction
becomes the transaction identifier (txid), which is included in the
Merkle tree that produces a block header and is ultimately recorded in
the block chain in an immutable way. Thus, the given inputs and outputs 
of a given transaction are provably linked together.

\subsubsection{Financial friction in Bitcoin transactions}

Miners are compensated by ``fees.'' The amount paid in fees for a given transaction 
is the difference between the combined values of the
inputs and the combined values of the outputs. Miners, who are limited in how many bytes they can fit a given block, generally choose to include the transactions with the most profitable fees 
with respect to the number of bytes in its raw transaction \cite[Chapter 5]{MasteringBitcoin}. 
When discussing our schema, we will denote the fees for a given transaction by $\fgeneric$. See Section \ref{cost} for estimations of these amounts. 

In order for a Bitcoin transaction to be considered valid it must satisfy certain basic properties such as not double spending a previously spent output, having valid signatures, etc. Any block that contains an invalid transaction will be rejected by the network. In addition, the Bitcoin Core software distributions to miners suggests requirements that transactions need to satisfy in order to be considered ``standard.'' These requirements are implemented at the discretion of each miner and thus vary slightly across the network; a miner may refuse to include a given transaction in the blocks he mines as ``non-standard,'' but if another miner broadcasts a block with this transaction in it, he will still accept that block if the transaction is valid. In particular, for a transaction to be considered standard, each of its non-$\opreturn$ output must have a minimal value so as to prevent the network from being spammed by extremely low value transactions. Any amount of bitcoin below this minimum is called ``dust.'' As of version 0.14 (March 2017), Bitcoin Core \cite{Dustcode} recommends that that miners refuse transactions that have a P2PKH output of less than $.00000546$ bitcoin, currently (June 2017, 1BTC=2720 USD) around $.01$ USD. We denote by $\mathcal{D}$ this minimal amount. Fees and the requirement to leave dust can greatly erode the value of a user's bitcoins if she engages in many transactions of small amounts.



\subsection{MIT Media Labs certificate issuing schema} 
We are inspired by the transaction structure used in \cite{MIT}. 
In this system a certificate or diploma is issued to a user who
completes a given program of study, encoded in a Bitcoin transaction. 
The transaction has a single input, from
the credential issuer, so the transaction must be signed by the
private key corresponding to the issuer's address. Hence, verifiers can be
confident that credential was issued by an approved party. There are three outputs. The first is the Bitcoin address of
the user. Then the user can
authenticate herself as the holder of the credential by signing
messages using the corresponding private key. The second output
is to an address again belonging to the issuer. If this output
is spent, the certificate is seen as being revoked. We view this revocation mechanism as a key innovation of \cite{MIT}, and we integrate and develop it into our system. 
Finally, the third
output is an $\opreturn$ that contains the certificate
information. Note that as each of these UXTOs is thought
of as having symbolic meaning, their bitcoin values are
secondary; indeed, they are assigned values slightly larger than $\mathcal{D}$.


\begin{figure}{
	\centering 
	\begin{tikzpicture}
	\coordinate (up_left) at (0,1.8);
	\coordinate (bottom_left) at (0,0);
	\coordinate (up_middle) at (5,1.8);
	\coordinate (bottom_middle) at (5,0);
	\coordinate (bottom_right) at (10.5,0);
	\coordinate (up_right) at (10.5,1.8);
	
	\draw (bottom_left) -- (bottom_right) ;
	\draw (bottom_right) -- (up_right) ;
	\draw (up_left) -- (up_right) ;
	\draw (up_left) -- (bottom_left) ;
	\draw (up_middle) -- (bottom_middle) ;
	
	\draw (up_left) node[above right] {Input Addresses};
	\draw (up_middle) node[above left] {Amounts};
	\draw (up_middle) node[above right] {Output Addresses};
	\draw (up_right) node[above left] {Amounts};
	\draw (up_left) node[below right] {$\begin{array}{l} \text{Issuer} \end{array}$};
	\draw (up_middle) node[below left] {$\begin{array}{r} .000155 \text{ BTC}\end{array}$};
	\draw (up_middle) node[below right] {$\begin{array}{l}\text{Recipient} \\\text{Issuer (for     revocation)}\\\opreturn( \text{ Certificate info})\end{array}$};
	\draw (up_right) node[below left] {$\begin{array}{r}.000275 \text{ BTC} \\ .000275 \text{ BTC} \\ 0 \text{ BTC}\end{array}$};
	\draw (bottom_middle) node [above right] {Fees:};
	\draw (bottom_right) node [above left]{$.0001$ BTC};
	
	\end{tikzpicture}
}
	\caption{Schema of an MIT certificate issuing transaction as in \cite{MIT}. See, for example, txid: \href{https://blockchain.info/tx/41740ae0812e5a7804778f43c9fd1f8df50fe1bcd0545e9d627a83ab9d0d3d07}{41740ae0812e5a7804778f43c9fd1f8df50fe1bcd0545e9d627a83ab9d0d3d07}}

	\label{fig:transactionMIT}

\end{figure}

\subsection{The DLREP function} \label{DLREP} 
In~\cite{Brands}, Brands proposed very efficient ways of
revealing parts of an identity to verifiers, relying on discrete
logarithms and hash functions. All the following is
from~\cite{Brands}. Assume that $n$ identity fields $X_1,\dots,X_n$ are to be
cryptographically blinded  for further proofs. Let $q$ be a prime
number and $G$ a group of order $q$, in which the discrete logarithm is
hard. Typically, we take $G$  to be the Koblitz elliptic curve secp256k1 where points are represented with 64 bytes (we use multiplicative notation for compatibility with \cite{Brands}), namely we use the same $G$ that is already being used for the Bitcoin signature protocol. Let $g_0,g_1,\ldots,g_n \in G$. Furthermore, there is the need (see Section \ref{Proposal}) for an
auxiliary random $X_0$ to protect unknown fields from a dictionary attack when the other fields are known.

\begin{defin}
  The tuple $(X_0,X_1,\ldots,X_n) \in \mathbb{Z}_q^{n+1}$ is called a
  Discrete Logarithm REPresentation (DLREP) of
  $h=\prod\limits_{j=0}^{n} g_j^{X_j} \in G$ with respect to
  $(g_0,g_1,\ldots,g_n)$.
\end{defin}

To (non-interactively) prove  knowledge of a DLREP of $h$ to a verifier $\Verifier$, a
prover $\Prover$ performs the following protocol
steps~\cite[\textsection 2.4.3]{Brands}
\begin{enumerate}
\item $\Prover$ generates $n+1$ secret, random numbers $a_0,a_1,\ldots,a_n$ in
  $G$. Let $A=\prod\limits_{j=0}^{n} g_j^{a_j}$, and compute $c$ as $c=\Hash(A)$, 
  where $H$ is a one-way
  hash function. 
\item $\Prover$ computes $b_j=a_j+c X_j$, $j=0,1,\ldots,n$ and sends them, as well as $c$ to $\Verifier$.
\item The verifier $\Verifier$  checks that
$\Hash(\prod\limits_{j=0}^{n} g_j^{b_j} h^{-c})=c$ holds.
\end{enumerate}
Then, \cite[Chapter 3]{Brands} shows how the DLREP can be used to selectively prove properties about the $X_j$'s, while any other information remains
hidden. These techniques can be used to prove arbitrary satisfiable Boolean statements about the
$X_j$'s. For example, a prover can demonstrate that she is a French citizen AND that she is either under 18 OR over 65. $\Prover$ can prove (true) statements about her identity that contain an arbitrary number of ANDs, ORs, and NOTs in such a way that $\Verifier$ only learns information that can be computed using the status of the formulas requested and information available a priori. See \cite[Proposition 3.6.1]{Brands} for a formal statement of this result. Brands~\cite{Brands} also shows that if the discrete logarithm problem is difficult, DLREP is one-way and collision-intractable, preventing an adversary from forging an identity with a given DLREP.

\section{Our proposal} \label{Proposal}
\subsection{Actors, protocol structure, and security assumptions}\label{subsec:entities}
Our system will have three types of actors :
{\bf Identity Providers} $(\IP)$,
{\bf Service Providers} $(\SP)$, and {\bf Users} $(\USR)$.
We borrow the following from~\cite{MIT}:
\begin{defin} An identity is a tuple
$(X_1, \ldots, X_n)$ where each $X_j\in \mathbb{Z}_q$ stands for a different
attribute, as exemplified below.
\end{defin}
An attribute $X_j$ may represent a name, a date of birth, a social insurance number, medical or financial data, or some other personal information about a user. Typically, based on an identity provided by $\IP$, a user ($\USR$)
wants to convince $\SP$ to give her access to its services.

\subsubsection{Assumptions on actors}  We
consider that both the Bitcoin addresses of $\IP$ and $\SP$ are
well-established and public, $\addr_{\IP}$ and $\addr_{\SP}$ respectively. We will consider scenarios in which we have multiple identity
providers and service providers, whose addresses are denoted $\addr_{\IP_1},  \addr_{\IP_2}, \ldots $ and $\addr_{\SP_1},  \addr_{\SP_2}, \ldots $ respectively. In contrast, $\USR$ may have
different Bitcoin addresses $\addr_{\USR}^{(1)}, \addr_{\USR}^{(2)}, \ldots $ in order to 
obfuscate the link 
between her identity transactions. When discussing a given user's address generally, we write $\addr_{\USR}^{(i)}$ to indicate one her addresses. Note that a user should not re-use Bitcoin addresses 
that she has used for non-identification transactions, in
order to not link this identity with her other Bitcoin 
activity.

We assume that all of $\USR$, $\IP$, and $\SP$ are capable of sending and receiving bitcoin and that they can perform operations in secp256k1. We will explore in Section \ref{real-world-security} further technical requirements on the ability of $\SP$ to track Bitcoin transactions which will depend on $\SP$'s security requirements. We assume that $\IP$ validates a user's real world identity (via a more or less rigorous verification process) and then publishes documents that are correct. Furthermore, $\IP$ should handle user personal data in a way that respects user-privacy. Note that $\IP$ does not need to stay online for the identities it issues to be used, and only participates for issuing and revocation of identities, and certain exceptional maintenance, see Section \ref{limiteduse}. Service providers accept identities issued by identity providers they wish to trust. Note that service providers may fail or refuse to provide a
service, a fact which can not be managed by our protocol. They may
deviate from the protocol (at the risk of impairing their reputation, see below).

\subsubsection{Assumptions on Bitcoin network} 

We will use the \emph{public ledger} functionality of Bitcoin: it
is a ``bulletin board'' where anyone can post messages and read messages
posted. More precisely, \cite{GarayKL15} and \cite{PassSS17}
provide the definitions of \emph{liveness}, i.e.\ every honest
participant will have its posted messages seen by every honest
participant after some delay, and \emph{persistence}, which means that
every posted message will indefinitely be seen at the same position by
all participants. We will also rely on the security semantics of the Bitcoin transaction verification procedure which ensure no
double-spending, that each non-generation transaction has inputs
linked to previous transaction outputs, etc. Under some quantitative bounds
on the relative power of the adversary, be it computing power
in~\cite{GarayKL15}, and or computing and network
power~\cite{PassSS17}, the Bitcoin core protocol is proven to securely
provide these functionalities.

The above results are theoretical and quantitative. There could be real
world situations in the Bitcoin blockchain  where the adversary has
enough power to violate the above quantitative bounds, and also
accidental cases where problems occurs like small forks, peer-to-peer
failures, etc. We will discuss the impact of these possible attacks
and failures in Section~\ref{real-world-security} below.

\begin{remark}
	
	Note that there are other relatively well-established blockchains such as Ethereum that can also serve as a ``bulletin board.'' However, by working in Bitcoin, we can use the linking mechanism of Bitcoin transactions, which is not natively present in the account based model of Ethereum \cite{Ethyellow}. Also, the total hash power of the Bitcoin network is substantially greater than that of Ethereum \cite{Ethcomp}, which can be seen as a sign that Bitcoin has a great resilience to $51\%$ attacks.  
	
	\end{remark} 

There are three steps for our protocol: a {\bf Setup phase}, an {\bf
  Enrollment phase}, and an {\bf Operational phase}.

\subsection{Setup phase} 
Each $\IP$ will choose some set of $g_0, g_1 \ldots, g_n \in G$ 
that will serve as the base for a DLREP function. These $g_j$ should be public and readily available. 
For example, $\IP$ could create a series of Bitcoin transactions
with inputs from his address in which the $g_j$ and the fields they represent 
are stored in $\opreturn$ outputs.


\subsection{Enrollment phase}

During the Enrollment phase, $\USR$ brings to $\IP$ the (physical,
biometric, etc) elements required to assert that her identity indeed
matches all the $X_j$'s. This can be as strong as a physical meeting,
in which the user shows a passport, or as light as an authentication
on a web server, depending on the policy of $\IP$. During this phase
$\USR$ should provide $\IP$ with a Bitcoin address $\addr_{\USR}^{(i)}$
that she controls and an 
element $g_0^{X_0}$ to protect against dictionary attacks, where 
$X_0$ is chosen at random by $\USR$ so that $\IP$ does not learn
it. Then, $\IP$ can form 
$h_{\addr_{\USR}^{(i)}}=g_0^{X_0} g_1^{X_1} \ldots g_n^{X_n},$ as in
Section \ref{DLREP}.


The Enrollment phase corresponds to a single Bitcoin transaction, $\txpublish$.  The primary purpose of this transaction is to record $h_{\addr_{\USR}^{(i)}}$ in the blockchain; however, we see that this transaction will include other structure.

 {\bf $\txpublish$ (Identity Establishment) :} $\IP$ sends amounts of bitcoin to two outputs. First a minimal amount of bitcoin $\mathcal{D}$ is sent to the user's address $\addr_\USR^{(i)}$; this ties the user's address to the identity. Also, $\IP$ sends bitcoin to a $1$ of $2$ P2SH multisig of $\addr^{(i)}_{\USR}$ and $\addr_{\IP}$, denoted $\msig1\_2(\addr_\USR^{(i)},\addr_{\IP})$, which we view as an \textbf{authentication token} that the user will spend upon using her identity. Moreover, either $\USR$ or $\IP$ can prevent further use of the token by $\USR$ by sending it to $\IP$ or even spending it to a random address. This should be seen as \textbf{revocation}. More precisely, when using her identity as described below in Section \ref{sec:operational}, $\USR$ will send transactions of a specific form that return bitcoin to the same multisg address of $\addr_{\IP}$ and $\addr_{\USR}$ leaving a transaction output for future authentications; if at any point $\USR$ or $\IP$ spend this output in a transaction that is not of the form of another authentication, then this transaction is a $\txrevoke$ and the identity is seen as revoked.  
 Finally, an
	$\opreturn$ contains  $h_{\addr_{\USR}^{(i)}}$.



The authentication token will be used in subsequent transactions; its amount $V$ will be calibrated to cover the costs of these transactions, see Section \ref{cost}.

Note that the  structure of $\txpublish$ is similar to that of the transactions in the architecture of \cite{MIT} as shown in
Figure \ref{fig:transactionMIT}. Now, revocation 
can be performed by both $\IP$ and by $\USR$ as both parties can destroy the authentication token via a $\txrevoke$. 


\begin{figure}{ \small 
	\centering 
	\begin{tikzpicture}
	\coordinate (up_left) at (-5,2);
	\coordinate (bottom_left) at (-5,0);
	\coordinate (up_middle) at (0,2);
	\coordinate (bottom_middle) at (0,0);
	\coordinate (bottom_right) at (5,0);
	\coordinate (up_right) at (5,2);
	\coordinate (label_left) at (-5,2.5);
	\coordinate (label_middle) at (0,2.5);
	\coordinate (label_right) at (5,2.5);

	\draw (bottom_left) -- (bottom_right) ;
	\draw (bottom_right) -- (up_right) ;
	\draw (up_left) -- (up_right) ;
	\draw (up_left) -- (bottom_left) ;
	\draw (up_middle) -- (bottom_middle) ;
	
	\draw (up_left) node[above] {$\txpublish$};
	\draw (label_left) node[right] {Input Addresses};
	\draw (label_middle) node[left] {Amounts};
	\draw (label_middle) node[right] {Output Addresses};
	\draw (label_right) node[left] {Amounts};
	\draw (up_left) node[below right] {$\begin{array}{l}\addr_{\IP}\end{array}$};
	\draw (up_middle) node[below left] {$\begin{array}{r}V+\mathcal{D}+\fpublish\end{array}$};
	\draw (up_middle) node[below right] {$\begin{array}{l} \addr_\USR^{(i)} \\ \msig1\_2(\addr_\USR^{(i)},\addr_{\IP})\\\opreturn\left(h_{\addr_{\USR}^{(i)}}\right)\end{array}$};
	\draw (up_right) node[below left] {$\begin{array}{r} \mathcal{D} \\V\\\end{array}$};
	\draw (bottom_middle) node [above right] {Fees:};
	\draw (bottom_right) node [above left]{$\fpublish$};
	
	\end{tikzpicture}
}
	\caption{Structure of $\txpublish$.}
	
	\label{fig:transaction0}

\end{figure}

\begin{remark}
	
	There are alternative zero-knowledge selective credential systems in addition to that of Brands \cite{Brands}. As discussed above, one advantage of using Brands' scheme is that its cryptographic primitives: discrete logarithms (in our case on secp256k1) and hash functions are also primitives of Bitcoin, so we minimize the number of cryptographic assumptions necessary. Also, the commitments of Brands are small enough (a compressed elliptic curve point of $33$ bytes) to fit in an $\opreturn$. In contrast this is not the case for example for the commitments of the Camenisch-Lysyanskaya scheme which produces commitments of $670$ bytes \cite[Table 2]{rapport:idemix-spec}.

\end{remark}

\begin{remark}
	
	One can imagine cases where a hostile or hacked $\IP$ uses the authentication token to obtain services acting as if it were the user, possibly with the aim of harming the user's reputation. However, when spending a multisig output, it is visible which of the public keys one is signing by \cite{MasteringBitcoin}, thus such an attack would be visible and, in fact, damage $\IP$'s reputation.
	
	\end{remark}

\subsection{Operational phase} \label{sec:operational}

The Operational phase is made up of two further Bitcoin transactions. 
We think of certain outputs as being distinguished (or colored with a transferable semantic meaning 
in the sense of Colored Coins \cite[\textsection 9.2]{bitcoin-princeton}, \cite{OpenAssets}), corresponding to the authentication token.
The flow of this token will chain the transactions together and ultimately to the creation of the identity in $\txpublish$. We suppose $\SP$ informs $\USR$ of what statement about her identity she needs to prove to authenticate. Then we have the Bitcoin transactions:

	
	 {\bf $\txrequest$ (Request for Service):}  $\USR$ creates a transaction where the input is the $\msig1\_2(\addr_\USR^{(i)},\addr_{\IP})$ from $\txpublish$.  One output is sent to $\addr_{\SP}$. One output is sent back to $\msig1\_2(\addr_\USR^{(i)},\addr_{\IP})$ and will serve as the authentication token for future transactions. $\USR$ proves to $\SP$ the required Boolean statement about
	the $X_j$'~s without revealing them as in Section \ref{DLREP} (see below for a discussion of how this proof is transmitted and stored). 
	
	 {\bf $\txaccept$ (Acknowledgment of the Identity by $\SP$):} Upon validating the proof of $\USR$, checking that the authentication token is the result of a series of $\txrequest$'s  each of whose input is the output of the previous chained back to a $\txpublish$, checking that $\txpublish$ was issued by a trusted $\IP$, and verifying that the multisig output of the most recent $\txrequest$ has not been spent (namely that there has not been a $\txrevoke$), $\SP$ accepts $\USR$'s authentication and uses its output from $\txrequest$ to send bitcoins to $\addr_{\IP}$. 

\begin{figure}{
	\small 
	\centering 
	\begin{tikzpicture}
	\coordinate (tx_label) at (-4.6,1.8);
	\coordinate (up_left) at (-5,1.8);
	\coordinate (bottom_left) at (-5,0);
	\coordinate (up_middle) at (-.6,1.8);
	\coordinate (bottom_middle) at (-.6,0);
	\coordinate (bottom_right) at (6.9,0);
	\coordinate (up_right) at (6.9,1.8);
	\coordinate (label_left) at (-5,2.3);
	\coordinate (label_middle) at (0,2.3);
	\coordinate (label_right) at (6.9,2.3);
	
	\draw (bottom_left) -- (bottom_right) ;
	\draw (bottom_right) -- (up_right) ;
	\draw (up_left) -- (up_right) ;
	\draw (up_left) -- (bottom_left) ;
	\draw (up_middle) -- (bottom_middle) ;
	
	\draw (tx_label) node[above] {$\txrequest$};
	\draw (label_left) node[right] {Input Addresses};
	\draw (label_middle) node[left] {Amounts};
	\draw (label_middle) node[right] {Output Addresses};
	\draw (label_right) node[left] {Amounts};	
	\draw (up_left) node[below right] {$\begin{array}{l} {\color{red} \msig1\_2(\addr_\USR^{(i)},\addr_{\IP})} \end{array}$};
	\draw (up_middle) node[below left] {$V$};
	\draw (up_middle) node[below right] {$\begin{array}{l}\addr_{\SP}\\{\color{red} \msig1\_2(\addr_\USR^{(i)},\addr_{\IP}) } \\\opreturn(\text{proof-ref}) \end{array}$};
	\draw (up_right) node[below left] {$\begin{array}{r}\faccept+\mathcal{D} \\V-(\frequest+\faccept+\mathcal{D})\end{array}$};
	\draw (bottom_middle) node [above right]{Fees:}; 
	\draw (bottom_right) node [above left]{$\frequest$}; 
	
	\coordinate (tx_label2) at (-4.6, -.5);
	\coordinate (up_left2) at (-5,-.5 );
	\coordinate (bottom_left2) at (-5,-1.6 );
	\coordinate (up_middle2) at (-.6,-.5 );
	\coordinate (bottom_middle2) at (-.6,-1.6 );
	\coordinate (bottom_right2) at (6.9,-1.6);
	\coordinate (up_right2) at (6.9,-.5);
	
	\draw (bottom_left2) -- (bottom_right2) ;
	\draw (bottom_right2) -- (up_right2) ;
	\draw (up_left2) -- (up_right2) ;
	\draw (up_left2) -- (bottom_left2) ;
	\draw (up_middle2) -- (bottom_middle2) ;
	
	\draw (tx_label2) node[above] {$\txaccept$\phantom{i}};
	\draw (up_left2) node[below right] {$\begin{array}{l}\addr_{\SP}\end{array}$};
	\draw (up_middle2) node[below left] {$\faccept+\mathcal{D}{}$};
	\draw (up_middle2) node[below right] {$\begin{array}{l}\addr_{\IP}\end{array}$};
	\draw (up_right2) node[below left] {$\begin{array}{r}\mathcal{D} \end{array}$};
	\draw (bottom_middle2) node [above right]{Fees:}; 
	\draw (bottom_right2) node [above left]{$\faccept$}; 
	\end{tikzpicture}
}
	\caption{The transactions that compose a typical authentication The inputs and outputs
		highlighted in red are thought of as an authentication token that chain the user's transactions together and to $\txpublish$.}
	\label{fig:transactionsopsalt}
\end{figure}

\subsubsection{Storage of proofs} A careful reading of \cite[Chapter 3]{Brands} shows that the size of the Brands proofs required to demonstrate a given Boolean statement about an identity $(X_0,\ldots, X_n)$ scales linearly in $n$, but also depends on the 
statement being proven. We note that these proofs will
generally be too large to 
be contained directly in an $\opreturn$. Depending on the needs of $\USR$ and $\SP$, we propose three different mechanisms by which these proofs might be transmitted and stored. \begin{inparaenum} \item A user can 
	store in the $\opreturn$ of $\txrequest$ a link to a site where the proofs are
	stored externally as well as a hash of the relevant contents of this site. We denote this information by proof-ref. The hash will be 
	included in mined blocks, so the information on the site has the same 
	protections against mutability as other information on the blockchain. 
	This 
	is similar to how metadata is stored in \cite{OpenAssets}. 
	  \item A user that is very concerned about privacy, or who is proving a statement that is already sensitive, can transmit the Brands proofs entirely off-chain. \item If one wants to avoid 
	  an off-chain storage mechanism, there are a number of non-$\opreturn$ ways to store data in the Bitcoin blockchain (see \cite{OpenAssets}) such as in a vanity address or using a fake $1$ of $N$ multisig. Alternatively, one can issue a P2SH output in $\txrequest$ with Pubkey Script $\texttt{OP\_{HASH160}}$ $H(\text{data})$ $\texttt{OP\_{EQUAL}}$ for which the corresponding input Sig Script is simply the data itself (see txid \href{https://blockchain.info/tx/db195e4bfcfb3cc6d47f8d6231cb59e543c31e01d196d557457bca0fa5c1aba0}{db195e4bfcfb3cc6d47f8d6231cb59e543c31e01d196d557457bca0fa5c1aba0}). While there are still limits on how much data can be placed in a single input, through using multiple inputs, one can store larger amounts of data in this fashion in exchange for paying (much) higher transaction fees.    
  \end{inparaenum} 

For the remainder of this article (and in Figure \ref{fig:transactionsopsalt}) we assume that proofs are being referenced via a link and a hash in an $\opreturn$.




\begin{remark} \label{r2} 
	$\txaccept$ publicly shows that $\SP$ has accepted
	$\USR$'s identity proofs as valid, contributing to the reputation of $\addr_{USR}^{(i)}$ (see Section \ref{repu}). This is particularly useful if the proofs
	were conveyed off-chain or are otherwise unavailable. $\txaccept$ can also serve to 
	alert $\IP$ that $\SP$ has used an identity that it
	provided, and can even be a basis for a payment by $\SP$ for the
	issuing of this identity.

\end{remark}



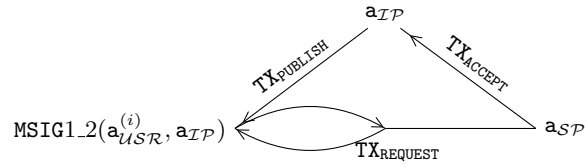
\begin{figure} \centering 
	\begin{tikzpicture}[decoration={
		markings,
		mark=
		at position 2cm
		with
		{
			\draw (-2pt,2pt) -- (2pt,0pt);
			\draw (-2pt,-2pt) -- (2pt,0pt);
		}
	}
	]*2/3
	\coordinate (bottom_left) at (-3*2/3,0);
	\coordinate (up_middlel) at (-.35*2/3,2*2/3);
	\coordinate (up_middler) at (.3*2/3,2*2/3);
		\coordinate (up_middle) at (0,2*2/3);
	\coordinate (bottom_right) at (3*2/3,0);
	\coordinate (bottom_middle) at (0,0);
		\coordinate (bottom_nearmiddle) at (.25*2/3,-.1*2/3);
	\coordinate (below_left) at (-3*2/3,-1*2/3);
	\coordinate (below_right) at (3*2/3,-1*2/3);
	
	
	\draw (up_middle) node[above] {$\addr_{\IP}$};
	\draw (bottom_left) node[left] {$\msig1\_2(\addr_\USR^{(i)},\addr_{\IP}) $};
	\draw (bottom_right) node[right] {$\addr_{\SP}$};
	\draw (bottom_nearmiddle) node[below] {$\txrequest$};
	\draw [postaction={decorate}] (up_middlel) -- (bottom_left) node[midway,above,sloped] {$\txpublish$} ;
	\draw [postaction={decorate}] (bottom_middle) -- (bottom_right) ;
	\draw [postaction={decorate}] (bottom_right) -- (up_middler) node[midway,above,sloped] {$\txaccept$} ;
	\draw [postaction={decorate}] (bottom_middle) to[bend left] (bottom_left) ;
	\draw [postaction={decorate}] (bottom_left) to[bend left] (bottom_middle) ;
	\end{tikzpicture}
	
	\caption{Scheme for users to prove their identity to service providers. 
		An identity is issued via a $\txpublish$. Subsequently, each
		transaction takes as input the output of a previous transaction ($\txpublish$ for the first authentication, $\txrequest$ thereafter). For simplicity, the $\opreturn$  output is not shown. 
	}
\end{figure}

\begin{remark} \label{remark: loopingmodels}
	
The transaction output of $\txrequest$ to $\msig1\_2(\addr_\USR^{(i)},\addr_{\IP}) $ is necessary if the user wishes to reuse an identity, so that $\USR$ will have an UTXO tied to her identity to spend in a subsequent $\txrequest$. This cyclic structure chains the various authentications together permitting a verifier to trace any of them back to the original identity issued in $\txpublish$. 
Alternatively, $\USR$ can obtain a new $\txpublish$ attached to a different address $\addr_{USR}^{(j)}$ for each authentication if she wishes to maintain a more complete anonymity in her authentications. However, doing so is slightly more expensive as additional fees need to be paid for each $\txpublish$. Additionally, this limits the ability of a user to take advantage of the reputation system we propose in Section \ref{repu}. (See the discussion in the introduction on user empowerment over her level of privacy and compare to \cite[Chapter 5.2.1]{Brands}, where Brands discusses the balance between reputation and anonymity and proposes reuse solutions for his certificates.) In cases where the identity is only designed to be used one time, this transaction output is unnecessary.

\end{remark}

\subsection{Limited use identities and setting Bitcoin values} \label{limiteduse}

Due to the chaining of authentications, when $\SP$ verifies the continued validity of $\USR$'s identity, the number of times this identity has been used can also be calculated. Thus, if $\IP$ includes a use limit of $N$ authentications with $h_{\addr_{USR}^{(i)}}$ in the $\opreturn$ of $\txpublish$, $\SP$ can check if this identity can still be used. Then, $\IP$ should calibrate the amount of bitcoin, $V$, that is placed in the $\addr_{\USR}^{(i)}$ output to cover these $N$ authentications. As we saw in Figure \ref{fig:transactionsopsalt} that each authentication consumes $\frequest+
\faccept+\mathcal{D}$ bitcoin before the authentication token is returned to the user, and this returned token needs to have a value of at least $\mathcal{D}$ after the last usage for the transaction to be accepted as standard, $V$ must be at least $N(\frequest+
\faccept+\mathcal{D})+\mathcal{D}$. Note, the fees required for a transaction to be processed in a timely fashion slowly vary based on market forces, so $\IP$ should, in practice, set $V$ to be slightly larger than current market demands in case miners increase their fees. Then, situations requiring $\IP$ to come online and top up its users' balances
can be limited to cases of extreme changes in Bitcoin fees.

\subsection{Coordinating multiple identities} \label{subsec:transactioncoordinate}

Suppose a given user has obtained identities $h_{\addr_{\USR}^{(1)}}$ and $h_{\addr_{\USR}^{(2)}}$ from more than one identity provider. We see that these identities can be coordinated.
	

\begin{figure}
	{\small
	\centering 
	\begin{tikzpicture}
	\coordinate (tx_label) at (-4.2,2.3);
	\coordinate (up_left) at (-5,2.3);
	\coordinate (bottom_left) at (-5,0);
	\coordinate (up_middle) at (-.6,2.3);
	\coordinate (bottom_middle) at (-.6,0);
	\coordinate (bottom_right) at (6.9,0);
	\coordinate (up_right) at (6.9,2.3);
	\coordinate (label_left) at (-5, 2.8);
	\coordinate (label_middle) at (-.6, 2.8);
	\coordinate (label_right) at (6.9, 2.8);

	\draw (bottom_left) -- (bottom_right) ;
	\draw (bottom_right) -- (up_right) ;
	\draw (up_left) -- (up_right) ;
	\draw (up_left) -- (bottom_left) ;
	\draw (up_middle) -- (bottom_middle) ;
	
	\draw (tx_label) node[above] {$\txrequestdouble$};
	\draw (label_left) node[right] {Input Addresses};
	\draw (label_middle) node[left] {Amounts};
	\draw (label_middle) node[right] {Output Addresses};
	\draw (label_right) node[left] {Amounts};	
	\draw (up_left) node[below right] {$\begin{array}{l} {\color{red} \msig1\_2(\addr_\USR^{(1)},\addr_{\IP})  } \\ {\color{blue} \msig1\_2(\addr_\USR^{(2)},\addr_{\IP})   }\end{array}$};
	\draw (up_middle) node[below left] {$\begin{array}{l} V_1 \\ V_2 \end{array}$};
	\draw (up_middle) node[below right] {$\begin{array}{l}\addr_{\SP}\\  {\color{red} \msig1\_2(\addr_\USR^{(1)},\addr_{\IP})   } \\ {\color{blue} \msig1\_2(\addr_\USR^{(2)},\addr_{\IP})   } \\\opreturn(\text{proof-ref}) \end{array}$};
	\draw (up_right) node[below left] {$\begin{array}{r}2\faccept+2\mathcal{D}  \\V_1-(\frequest+\faccept+\mathcal{D}) \\ V_2-(\frequest+\faccept+\mathcal{D}) \end{array}$};
	\draw (bottom_middle) node [above right]{Fees:}; 
	\draw (bottom_right) node [above left]{$\frequestdouble$}; 
	\coordinate (tx_label) at (-4.2, -.5);
	\coordinate (up_left) at (-5,-.5 );
	\coordinate (bottom_left) at (-5,-2 );
	\coordinate (up_middle) at (-.6,-.5 );
	\coordinate (bottom_middle) at (-.6,-2 );
	\coordinate (bottom_right) at (6.9,-2 );
	\coordinate (up_right) at (6.9,-.5 );
	
	\draw (bottom_left) -- (bottom_right) ;
	\draw (bottom_right) -- (up_right) ;
	\draw (up_left) -- (up_right) ;
	\draw (up_left) -- (bottom_left) ;
	\draw (up_middle) -- (bottom_middle) ;
	
	\draw (tx_label) node[above] {$\txacceptdouble$\phantom{i}};
	\draw (up_left) node[below right] {$\begin{array}{l}\addr_{\SP} \end{array}$};
	\draw (up_middle) node[below left] {$\begin{array}{l} 2\faccept+2\mathcal{D}   \end{array} $};
	\draw (up_middle) node[below right] {$\begin{array}{l}\addr_{\IP_1} \\ \addr_{\IP_2}  \end{array}$};
	\draw (up_right) node[below left] {$\begin{array}{r}\mathcal{D} \\ \mathcal{D}  \end{array}$};
	\draw (bottom_middle) node [above right]{Fees:}; 
	\draw (bottom_right) node [above left]{$\facceptdouble$}; 
	\end{tikzpicture}
	
}
		\caption{Use of the authentication tokens of two identities together. The paths of these tokens are colored in red and blue. The Brands proofs referenced in proof-ref are with respect to the DLREP of equation \ref{eqn:coord}}
	\label{fig:transactioncoordinate}
\end{figure}

Concretely, suppose a user has been issued an identity by $\IP_1$ consisting of $h_{\addr_{\USR}^{(1)}}=\prod\limits_{j=0}^{n} g_j^{X_j}$ and another identity by $\IP_2$ consisting of $h_{\addr_{\USR}^{(2)}}=\prod\limits_{u=0}^{m} (g'_u)^{Y_u}$. A service provider will be able to verify that each of these values correspond to the respective $\txpublish$ transactions issued by the identity providers. Then \begin{equation} h_{\addr_{\USR}^{(1)}}\cdot h_{\addr_{\USR}^{(2)}}=\prod\limits_{j=0}^{n} g_j^{X_j} \prod\limits_{u=0}^{m} (g'_u)^{Y_u} \label{eqn:coord} \end{equation}  is a 
DLREP commitment of the union of $X_0, \ldots X_n$ and $Y_0,\ldots, Y_m$. The user can do proofs with selective disclosure using this commitment. 


We preserve the chaining properties by having two transactions $\txrequestdouble$, which takes in the authentication tokens from both identities, and $\txacceptdouble$, which notifies both identity providers. The amounts used in these transactions are chosen as in Figure \ref{fig:transactioncoordinate} to ensure that a user's balances decrease by no more than what would have been the case for separate authentications with the two identities, in keeping with the calibration of $V$ in Section \ref{limiteduse}. (We will see in Section \ref{cost}, $\frequestdouble\leq 2\frequest$ and $\facceptdouble\leq 2\faccept$, so adequate fees are paid here; the change can be split between $\USR$'s authentication tokens for use in case of future Bitcoin fee increases, paid to $\SP$, or left to the miners to increase the speed of the transaction's approvals). 
This schema can obviously generalize to more than two identities.


Thus, a user can obtain many ``micro-identities'' - from the government, from her bank, from her employer, from her health care provider - which she can manage together without having to unnecessarily share information between her identity providers. This is very much in the spirit of a PIMS \cite{PIMS}. 

\begin{remark} \label{remark:coordination} The ability of $\USR$ to issue a transaction as in Figure \ref{fig:transactioncoordinate}, which requires signing with the private key corresponding to the address of each identity, is already a weak way of establishing that these identities belong to the same person. However, it is possible for malicious users to pool the private keys from identities corresponding to distinct people. $\USR$ can provide stronger proof of the connection of her identities if she shows as part of her proof in $\txrequest$ that $h$ and $h'$ share common fields, such as name or social insurance number. 
\end{remark} 

	
	



\subsection{Estimates of cost}\label{cost}


We now estimate the costs of the transactions we have introduced in the preceeding sections. As mentioned in Section \ref{sec:BuildingBlocks}, Bitcoin miners have flexibility in what fees they demand. However, the current standard fee to have one's transaction processed in a timely manner is 360 satoshis, namely $.0000036$ bitcoins, per byte \cite{fees}.  Based on our schema, $\txpublish$ will contain one input, one P2PKH output, one P2SH output, and an $\opreturn$ that contains one (compressed) point on secp256k1. Hence the $\opreturn$ contains $33$ bytes resulting in a total transaction size of roughly $267$ bytes (see \cite[Chapter 2]{MasteringBitcoin} for more information on the size of the various components of a Bitcoin transaction) costing $.0009612$ bitcoin. At current market rates (June 2017, 1BTC=2720 USD), this corresponds to a minimum transaction fee of approximately $ 2.61$ USD. 
We compute the sizes and costs of the other transactions similarly (based on proof-ref consisting of a $32$ byte SHA-256 hash and a $30$ byte url when necessary):  \begin{center} \begin{tabular}{|c| cccccc| }
	  \hline
	Transaction & $\txpublish$ & $\txrevoke$ & $\txrequest$ & $\txaccept$ & $\txrequestdouble$ & $\txacceptdouble$ \\ 
	  \hline
	$\#$ Bytes & $267$ & $229$ & $334$ & $191$ & $479$ & $225$ \\
	  \hline
	Cost (USD) & $2.61$ & $2.24$ & $3.28$ & $1.87$ & $5.41$ & $2.20$ \\
	  \hline
\end{tabular} \end{center}



Then, building off Section \ref{limiteduse}, the total cost to issue an $N$ use identity is the value of the input issued by $\IP$ in $\txpublish$. As in Figure \ref{fig:transaction0}, this is  
$$\begin{array}{l c l} \text{Cost of }N\text{-use Id} &= &V+\mathcal{D}+\fpublish \\  & = & N(\frequest+\faccept+\mathcal{D})+2\mathcal{D}+\fpublish \\ & \approx & 5.2N+2.6 \text{ USD}. \end{array}$$ 

Note that Bitcoin fees have increased substantially recently as the Bitcoin community seeks consensus on how to scale block capacity. It is hoped that a solution to this issue, such as an implementation of SegWit, will reduce fees \cite{Segwitonfees}.



\subsection{Obtaining information about the Bitcoin network} \label{obtain-info}

Note that in the processing of an authentication, it is the service provider that must verify the status of past Bitcoin transactions.
Service providers with rigorous verification requirements, such as banks and insurance companies, should run a full node or possibly a Simplified Payment Verification (SPV) client, see \cite{MasteringBitcoin}. Note the SPV protocol, which is already commonly used by vendors who payment in Bitcoin, allows someone who downloads merely the $80$ byte header of each block to verify that a given transaction has been included in a block, upon being provided with information related to that transaction by a full node. Hence, a service provider running this protocol can verify that each of the $\txrequest$'s a user has issued, chained back to $\txpublish$, counting the number of times the identity has been used. This process also checks that the identity has not been revoked as the SPV client sees that the network has accepted the most recent transaction, so the transaction output controlled by the multisig between $\USR$ and $\IP$ could not have already been spent in a $\txrevoke$. Service providers with less rigorous standards may retrieve their information from an online block explorer if they accept the additional risks of attacks on these sites.

\subsection{Security considerations in case of blockchain failures}\label{real-world-security}


In Section \ref{subsec:entities}, we place ourselves in a security model in which Bitcoin possesses certain properties of an ideal blockchain. Here we explore the consequences on our system when these properties are not satisfied. 

\textbf{Inconsistencies in the Bitcoin ledger:} The integrity of the Bitcoin ledger serves in our system to allow issuer oversight, concretely to allow the issuer to revoke identities and to impose limits on the number of uses. On the other hand, if there is a fork, a dishonest user can to continue to use an identity which an issuer has revoked until the revocation transaction finally appears in the dominant chain. If an attacker can issue a double spend (due to an accidental fork, because the attacker has a large percentage of the mining power, etc), then she can reuse her authentication token allowing her to exceed her usage limit. 

\textbf{Bitcoin network failure:} We also rely on Bitcoin P2P infrastructure to propagate the transactions that make up our protocol, and we rely on being able to download information on previous Bitcoin transactions from nodes to check the state of an identity. An attack on the P2P Bitcoin network can translate into a denial of service attack on our system as one cannot issue $\txpublish$, $\txrequest$, etc if the network does not relay them or if one cannot verify relevant previous transactions. 
How vulnerable a service provider is to network attacks will depend on how it receives information about the network as in Section \ref{obtain-info}. Note that, regardless of this choice, user privacy is protected and impersonation is prevented 
by the security of Brands' protocols, see \cite{Brands}. Even a service provider that obtains its information from a block explorer can assure itself of the correctness of Brands proofs and the validity of signatures.


\section{Example use cases} \label{sec:examples}

In this section we propose a few use cases of our system that highlight its advantages versus existing systems.

\subsection{University ID} 

We consider a university where the
administration delivers identity credentials to students, teachers, and staff. These credentials
provide certificates of various fields related to the user including their name, their status at the university (student, teacher, etc), and their academic records. Individuals may use such identities, revealing some (or none) of these fields, to authenticate themselves to various university services such as the university pool or medical clinic. 

Now imagine that a user wants to claim a discount on car insurance reserved for students with high GPAs. This student may need to coordinate her university identity with a driver's license issued by her local government. Then she can selectively reveal information to the service provider, the insurer, using the multi-$\IP$ protocols described in Section \ref{subsec:transactioncoordinate}. If her status at the university changes, her university identity can be revoked preventing her from performing such authentications, even as her driver's license identity remains valid. 





\subsection{Network of small museums} \label{ex:museums}

We imagine a group of small museums that form a partnership in which any member of one museum is allowed a limited number of visits to the other museums. In this case, the user is a member of one of the museums, the identity provider is the museum that issued the membership, and the other museums are service providers. Then the user may selectively disclose fields such as her membership status or category of membership. More sensitive information may be included in the identity allowing the user to authenticate to tax authorities which give a tax credit for museum memberships. The limit on the number of visits is controlled through the methods of Section \ref{limiteduse}.



In contrast to the tax authorities, the security requirements of the museums may allow them to obtain the transaction information from an online block explorer, completely outsourcing the costs of transmitting and storing information to the Bitcoin network similar to how \cite{CryptID} uses the blockchain as a virtual server. This may be substantially cheaper and more streamlined than traditional systems (namely, either for each of the museums to invest in infrastructure that then has to be coordinated or for a single museum to set up infrastructure to manage the entire system which may create conflicts of interest and be unacceptable to the other museums). Thus, our system allows the museums to create a shared, neutral management space, maintaining transparency into exactly how the data is stored and used, that minimizes infrastructure costs. 



\section{Building a reputation on the blockchain} \label{repu} 

We see in Section \ref{ex:museums}, in the case of our museums, that little infrastructure is required of $\SP$. Nonetheless, $\SP$ must be able to compute in secp256k1, perform Bitcoin transactions, and be able to access the blockchain history, as discussed in Sections \ref{subsec:entities} and \ref{real-world-security}.
Imagine that some very lightweight service provider wants to participate in this network, but
does not have the security requirements, nor the resources to justify performing these operations. For example,
this may be the case of a university pool in the university ID example of Section \ref{sec:examples}. 

As all transactions are visible in the blockchain, a user can then simply direct 
a lightweight service provider to her past transactions, which requires merely an Internet connection, and prove that she controls the private key corresponding to those transactions by issuing a signature. Then, if the lightweight service
provider is willing to trust the larger service providers that have already accepted the
user's identity (e.g. if the university pool is willing to trust the campus medical clinic 
in accepting that the user is a member of the university community), it is not necessary to 
re-validate the relevant Brands proofs.  
As seen before (see Section \ref{subsec:transactioncoordinate}), 
a user may have had her identity established
under different Bitcoin addresses and proven to different service
providers in such a way that is unknown that these
addresses belong to the same user. 
If the user has used the two identities together in a $\txrequestdouble$, the light service provider may be again willing to trust that the other service provider has verified these two identities as corresponding to the same person.
Alternatively, in situations with lower security standards (as per Remark \ref{remark:coordination}), the user can issue signatures for both of the private keys corresponding to the identities used.

Moreover, the collection of transactions of a user, seen as having been accepted via $\txaccept$ transactions, gradually forms a digital footprint of the user. While some users will want to avoid reusing the same $\txpublish$ for multiple authentications for greater anonymity, for other users this digital presence, over which the user has a great deal of direct control, can be a useful addition to the online reputation they develop, for example through social media. 

\section{Conclusion}\label{Conclu}

%
	
The Bitcoin blockchain is a global network, and by building on top of this network, we can take advantage of its existing infrastructure
to reach a global scope while minimizing overhead. Moreover, by placing an identity management system in 
this decentralized space, we have seen that we can strike a more equitable 
balance between the rights and responsibilities of users and identity issuers.

	

\subsubsection*{Acknowledgment}
The work leading to this paper has received funding from the European Community's Framework Programme (FP7/2007-
2013) under Grant Agreement n° 607049.

\nocite{*}

\bibliographystyle{plain}

\bibliography{accgl-cbt.bib}

\begin{thebibliography}{10}

\bibitem{CryptID}
Crypt{I}{D}.
\newblock online, source code available at
  \url{https://github.com/CryptidID/Cryptid}, Consulted 2017 April.
\newblock \url{http://cryptid.xyz/}.

\bibitem{IDCoins}
I{D}{C}oins, Consulted 2017 April.
\newblock \url{https://github.com/IDCoin/IDCoin}.

\bibitem{fees}
Predicting {B}itcoin fees for transactions, Consulted 2017 April.
\newblock \url{https://bitcoinfees.21.co/}.

\bibitem{eestonia}
Estonian e-residency, Consulted 2017 March.
\newblock \url{https://e-estonia.com/e-residents/about/}.

\bibitem{PIMS}
Serge Abiteboul, Benjamin Andr{\'e}, and Daniel Kaplan.
\newblock Managing your digital life.
\newblock {\em Commun. ACM}, 58(5):32--35, April 2015.

\bibitem{Blockstack}
Muneeb Ali, Jude Nelson, Ryan Shea, and Michael~J. Freedman.
\newblock Blockstack: A global naming and storage system secured by
  blockchains.
\newblock In {\em 2016 USENIX Annual Technical Conference (USENIX ATC '16),
  Denver, CO, USA, June 22-24, 2016. Proceedings}, pages 181--194, 2016.

\bibitem{MasteringBitcoin}
Andreas~M. Antonopoulos.
\newblock {\em Mastering Bitcoin}.
\newblock O'Reilly Media, Sebastopol, California, 2015.
\newblock {ISBN:} 978-1-449-37404-4.

\bibitem{Brands}
Stefan Brands.
\newblock {\em Rethinking Public Key Infrastructures and Digital Certificates
  (Building in Privacy)}.
\newblock MIT Press, Cambridge, MA, USA, 2000.

\bibitem{DBLP:journals/ieeesp/CamenischLN12}
Jan Camenisch, Anja Lehmann, and Gregory Neven.
\newblock Electronic identities need private credentials.
\newblock {\em {IEEE} Security {\&} Privacy}, 10(1):80--83, 2012.

\bibitem{OpenAssets}
Flavien Charlon.
\newblock Open assets protocol (oap/1.0).
\newblock Online,
  \url{https://github.com/OpenAssets/open-assets-protocol/blob/master/specification.mediawiki},
  2011.

\bibitem{JH}
Ian~Miers Christina~Garman, Matthew~Green.
\newblock Accountable privacy for decentralized anonymous payments.
\newblock In {\em Financial Cryptography and Data Security}, 2016.

\bibitem{Dustcode}
The Bitcoin~Core developers.
\newblock Bitcoin transactions primitives code, Consulted 2017 March.
\newblock
  \url{https://github.com/bitcoin/bitcoin/blob/0.14/src/primitives/transaction.h}.

\bibitem{GarayKL15}
Juan~A. Garay, Aggelos Kiayias, and Nikos Leonardos.
\newblock The bitcoin backbone protocol: Analysis and applications.
\newblock In Elisabeth Oswald and Marc Fischlin, editors, {\em Advances in
  Cryptology - {EUROCRYPT} 2015 - 34th Annual International Conference on the
  Theory and Applications of Cryptographic Techniques, Sofia, Bulgaria, April
  26-30, 2015, Proceedings, Part {II}}, volume 9057 of {\em Lecture Notes in
  Computer Science}, pages 281--310. Springer, 2015.

\bibitem{ChainAnchor}
Thomas Hardjono, Ned Smith, and Alex~(Sandy) Pentland.
\newblock Anonymous identities for permissioned blockchains.
\newblock \url{http://www.the-blockchain.com/docs/MIT-ChainAnchor-DRAFT.pdf},
  January 2016.

\bibitem{Ethcomp}
Steven Hay.
\newblock Bitcoin vs ethereum: Cryptocurrency comparison, 2017 March.
\newblock
  \url{https://99bitcoins.com/bitcoin-vs-ethereum-cryptocurrency-comparison/}.

\bibitem{BlockStartupSurvey}
Ori Jacobovitz.
\newblock Blockchain for identity management.
\newblock
  \url{https://www.cs.bgu.ac.il/%7Efrankel/TechnicalReports/2016/16-02.pdf},
  December 2016.

\bibitem{revocationlists}
Yabing Liu, Will Tome, Liang Zhang, David~R. Choffnes, Dave Levin, Bruce~M.
  Maggs, Alan Mislove, Aaron Schulman, and Christo Wilson.
\newblock An end-to-end measurement of certificate revocation in the web's
  {PKI}.
\newblock In {\em Proceedings of the 2015 {ACM} Internet Measurement
  Conference, {IMC} 2015, Tokyo, Japan, October 28-30, 2015}, pages 183--196,
  2015.

\bibitem{satoshi}
Satoshi Nakamoto.
\newblock Bitcoin: A peer-to-peer electronic cash system.
\newblock Online, \url{http://bitcoin.org/bitcoin.pdf}, 2008.

\bibitem{bitcoin-princeton}
Arvind Narayanan, Joseph Bonneau, Edward Felten, Andrew Miller, and Steven
  Goldfeder.
\newblock {\em Bitcoin and Cryptocurrency Technologies: A Comprehensive
  Introduction}.
\newblock Princeton University Press, 2016.

\bibitem{MIT}
Juliana Nazar{\'e}, Kim Hamilton, and Philipp Schmidt.
\newblock Digital certificates project.
\newblock online, source code available at
  \url{https://github.com/digital-certificates}, Consulted 2016 December.
\newblock \url{http://certificates.media.mit.edu}.

\bibitem{CanadaPrivacy}
{Office of the Privacy Commissionner of Canada}.
\newblock Privacy and your reputation - who shapes your identity online?, 2012.

\bibitem{PassSS17}
Rafael Pass, Lior Seeman, and Abhi Shelat.
\newblock Analysis of the blockchain protocol in asynchronous networks.
\newblock In Jean{-}S{\'{e}}bastien Coron and Jesper~Buus Nielsen, editors,
  {\em Advances in Cryptology - {EUROCRYPT} 2017 - 36th Annual International
  Conference on the Theory and Applications of Cryptographic Techniques, Paris,
  France, April 30 - May 4, 2017, Proceedings, Part {II}}, volume 10211 of {\em
  Lecture Notes in Computer Science}, pages 643--673. Springer, 2017.

\bibitem{estoniannews}
Giulio Prisco.
\newblock Estonian government partnerts with bitnation to offer blockchain
  notarization services to e-residents, 2015 November.
\newblock
  \url{https://bitcoinmagazine.com/articles/estonian-government-partners-with-bitnation-to-offer-blockchain-notarization-services-to-e-residents-1448915243/}.

\bibitem{rapport:idemix-spec}
Security Team.
\newblock Specification of the identity mixer cryptographic library version
  2.3.0.
\newblock Technical Report RZ 3730, IBM Research, Computer Science Dept, IBM
  Research -- Zurich, Switzerland, 2010.
\newblock 48 pages.

\bibitem{Segwitonfees}
Kyle Torpey.
\newblock Are bitcoin miners making more money off small blocks?, 2017 March.
\newblock
  \url{https://bitcoinmagazine.com/articles/are-bitcoin-miners-making-more-money-small-blocks/}.

\bibitem{Ethyellow}
Gavin Wood.
\newblock Ethereum: A secure decentralised generalised transaction ledger,
  {EIP}-150 REVISION (030c1b5 - 2017-07-10).
\newblock \url{https://ethereum.github.io/yellowpaper/paper.pdf}.

\bibitem{BlockIDSurvey}
Danny Yang, Jack Gavigan, and Zooko Wilcox-O'Hearn.
\newblock Survey of confidentiality and privacy preserving technologies for
  blockchains.
\newblock
  \url{https://z.cash/static/R3_Confidentiality_and_Privacy_Report.pdf},
  November 2016.

\bibitem{Zyskind}
Guy Zyskind, Oz~Nathan, and Alex Pentland.
\newblock Decentralizing privacy: Using blockchain to protect personal data.
\newblock In {\em 2015 {IEEE} Symposium on Security and Privacy Workshops,
  {SPW} 2015, San Jose, CA, USA, May 21-22, 2015}, pages 180--184, Los
  Alamitos, CA, USA, 2015. IEEE Computer Society.

\end{thebibliography}

\end{document}